\documentclass [dvips,12pt]{article}

\usepackage{bbm,latexsym,epsfig}

\newcommand{\be}{\begin{equation}}
\newcommand{\ee}{\end{equation}}
\newcommand{\ba}{\begin{eqnarray}}
\newcommand{\ea}{\end{eqnarray}}
\newcommand{\no}{\nonumber\\}

\newcommand{\lesssim}{ \ \mbox{\raisebox{-3pt}{$\stackrel%
{\displaystyle <}{\sim}$}} \ }
\newcommand{\gtrsim}{\:\mbox{\raisebox{-3pt}{$\stackrel%
{\displaystyle >}{\sim}$}}\:}

\newcommand{\mnu}{\mathcal{M}_\nu}
\newcommand{\deltasol}{\Delta m^2_\odot}

\begin{document}

\title{
\normalsize \hfill UWThPh-2009-16 \\
\normalsize \hfill CFTP/09-041 \\[8mm]
\LARGE
TeV-scale seesaw mechanism \\
catalyzed by the electron mass
}

\author{
W.~Grimus$\, ^{(1)}$\thanks{E-mail: walter.grimus@univie.ac.at}
\ and
L.~Lavoura$\, ^{(2)}$\thanks{E-mail: balio@cftp.ist.utl.pt}
\\*[3mm]
\small $^{(1)}$ University of Vienna, Faculty of Physics \\
\small Boltzmanngasse 5, A--1090 Vienna, Austria
\\*[2mm]
\small $^{(2)}$ Technical University of Lisbon \\
\small Centre for Theoretical Particle Physics, 1049-001 Lisbon, Portugal
}

\date{31 March 2010}

\maketitle

\begin{abstract}
We construct a model in which the neutrino Dirac mass terms
are of order the electron mass and
the seesaw mechanism proceeds 
via right-handed neutrinos with masses of order TeV.
In our model the spectra of the three light
and of the three heavy neutrinos
are closely related.
Since the mixing between light and heavy neutrinos is small,
the model predicts no effects in $pp$ and $p\bar p$ colliders.
Possible signatures of the model are
the lepton-number-violating process $e^- e^- \to H^- H^-$,
where $H^-$ is a charged scalar particle,
lepton-flavour-violating decays like $\mu^- \to e^- e^- e^+$,
or a sizable contribution
to the anomalous magnetic dipole moment of the muon. 
\end{abstract}

\newpage

\section{Introduction}

The discovery of neutrino oscillations
has established the existence of non-zero neutrino masses
in the sub-eV range.\footnote{Cosmological arguments~\cite{cosmology}
and the negative result of the direct search for neutrino mass
in tritium $\beta$ decay~\cite{tritium}
are also crucial in eliminating the possibility that
the neutrino masses may be higher than about one eV.
For a review on neutrino masses see~\cite{review}.}
An appealing way of explaining such small neutrino masses
is the seesaw mechanism~\cite{seesaw},
in which gauge-singlet right-handed neutrino fields $\nu_R$
with Majorana mass terms are added to the Standard Model (SM).
Those mass terms are not generated by the Higgs mechanism and may,
therefore,
be much larger than the electroweak scale.
Let $\nu_L$ be the neutral members
of the left-handed leptonic SM gauge 
doublets and $M_D$ and $M_R$ the mass matrices
of the fermionic bilinears $\bar \nu_R \nu_L$
and $\bar \nu_R C \bar \nu_R^T$,
respectively.
($C$ is the charge-conjugation matrix in Dirac space.)
We denote the mass scales of $M_D$ and $M_R$ by $m_D$ and $m_R$,
respectively.
If $m_R$ is much larger than $m_D$,
then an effective Majorana mass matrix
\be
\mnu = - M_D^T M_R^{-1} M_D
\label{mnu}
\ee
is generated for the $\nu_L$.
According to~(\ref{mnu}),
the scale $m_\nu$ of $\mnu$ is related to $m_D$ and $m_R$
through $m_\nu \sim \left. m_D^2 \, \right/ m_R$.
The mixing between the light and heavy neutrinos
is of order $m_D / m_R$,
hence very small.

We experimentally know that $m_\nu$
is in the range of $0.1\, \mathrm{eV}$ to $1\, \mathrm{eV}$,
if $m_\nu$ indicates the order of magnitude
of the largest of the light-neutrino masses, 
but,
in the framework of the seesaw mechanism,
$m_D$ and $m_R$ remain a mystery.
Since $M_D$ is the neutrino counterpart of the charged-fermion mass matrices,
$m_D$ may vary in between $100\, \mathrm{GeV}$
(which is both the electroweak scale and the top-quark mass scale)
and $1\, \mathrm{MeV}$
(the scale of the electron mass and of the up- and down-quark masses),
and this spans a range of five orders of magnitude.
As $m_R \sim \left. m_D^2 \, \right/ m_\nu$,
to these five orders of magnitude in $m_D$
correspond ten orders of magnitude in $m_R$.
If $m_D \sim 100\, \mathrm{GeV}$ then $m_R \sim 10^{13}\, \mathrm{GeV}$;
this is definitely below
the typical GUT scale $10^{16}\, \mathrm{GeV}$---identifying $m_R$
with the GUT scale would make the neutrino masses too small.
If instead $m_D \sim m_\tau$,
the mass of the tau lepton,
then $m_R \sim 10^9\, \mathrm{GeV}$.
If one wants to incorporate leptogenesis~\cite{leptogenesis}
into the seesaw mechanism,
then the appropriate $m_R$ would rather be $10^{11}\, \mathrm{GeV}$,
which lies in between the two previous estimates.
Finally,
we may as well use $m_D \sim m_e$,
the electron mass,
and then $m_R \sim 1\, \mathrm{TeV}$---this was noticed,
for instance,
in~\cite{soni1}.\footnote{Of course,
the simplest way to obtain $m_D \sim m_e$
is to \emph{assume} tiny Yukawa couplings,
as was done for instance in~\cite{aranda};
this is the opposite of what we do in this Letter---we discuss a scenario
with Yukawa couplings of order $0.01$--1.
Another avenue which has been pursued are technicolor models
with a suppressed $m_D$ and $m_R \lesssim 10^3\, \mathrm{TeV}$~\cite{shrock}; 
however,
in those models $m_D$ is not claimed to be as low as $m_e$.}
Such a low $m_R$ has the advantages that
it coincides with the expected onset of physics beyond the SM
and that it might produce testable effects of the seesaw mechanism
at either present or future colliders. 

The possibility that $m_D \sim m_e$ and
$m_R \sim 1\, \mathrm{TeV}$ is the starting point of this Letter.
We construct a seesaw model in which the vacuum expectation value (VEV)
responsible for the mass matrix $M_D$ is of order $m_e$.
In our model,
which is inspired by~\cite{HPSmodel} and~\cite{3p},
each charged lepton $\alpha$
($\alpha = e, \mu, \tau$)
has its own Higgs doublet $\phi_\alpha$,
whose VEV generates the mass $m_\alpha$.
On the other hand,
there is only one Higgs doublet $\phi_0$
which has Yukawa couplings to the $\nu_R$ and is,
therefore,
responsible for $M_D$.
We furthermore make use of a mechanism,
first put forward in~\cite{ma} and later extended in~\cite{nested},
for suppressing the VEV of $\phi_0$:
this doublet has a \emph{positive} mass-squared $\mu_0$
(in the scalar-potential term $\mu_0 \phi_0^\dagger \phi_0$)
and its VEV is triggered by a term $\phi_e^\dagger \phi_0$
in the scalar potential.
Since $\phi_e$ is the Higgs doublet which gives mass to the electron,
it must have a very small VEV,
and this explains the smallness of the VEV of $\phi_0$.

\section{The model}
\label{model}

\paragraph{Multiplets:} The gauge-$SU(2)$ multiplets of our model
are the following:
\begin{itemize}
\item Left-handed lepton doublets
$D_{L \alpha} = \left( \nu_{L \alpha},\, \alpha_L \right)^T$
and right-handed singlets $\nu_{R \alpha}$,
$\alpha_R$
($\alpha = e, \mu, \tau$).
\item Four Higgs doublets $\phi_0$ and $\phi_\alpha$.
\end{itemize}
We shall use indices $k$,
$l$ running over the range $0, e, \mu, \tau$.
The VEV of the neutral component of $\phi_k$ is denoted $v_k$.

\paragraph{Symmetries:} The family symmetries of our model
are the following:
\begin{itemize}
\item $\mathbbm{Z}_3$ symmetry $e \to \mu \to \tau \to e$.\footnote{If
we go beyond our main aim of achieving a light seesaw scale
and furthermore want to obtain the predictions
of maximal atmospheric-neutrino mixing
and a vanishing reactor-neutrino mixing angle,
then one may extend this $\mathbbm{Z}_3$ symmetry
to the full $S_3$ permutation symmetry of $e$,
$\mu$ and $\tau$~\cite{mu-tau}.}
\item Three $\mathbbm{Z}_2$ symmetries~\cite{HPSmodel}
\be
\mathbf{z}_\alpha: \quad
D_{L\alpha} \to - D_{L\alpha}, \ 
\alpha_R \to - \alpha_R, \
\nu_{R\alpha} \to - \nu_{R\alpha}.
\ee
Notice that
$\phi_\alpha$ does \emph{not} change sign under $\mathbf{z}_\alpha$.
The symmetries $\mathbf{z}_\alpha$
may be interpreted as discrete lepton numbers.
\item Three $\mathbbm{Z}_2$ symmetries~\cite{HPSmodel}
\be
\mathbf{z}^h_\alpha: \quad
\alpha_R \to -\alpha_R, \ \phi_\alpha \to -\phi_\alpha.
\ee
The $\mathbf{z}^h_\alpha$ are meant to ensure that $\alpha_R$
has no Yukawa couplings to the $\phi_\beta$ ($\beta \neq \alpha$).
\end{itemize}

\paragraph{Yukawa Lagrangian:}
The multiplets and symmetries of the theory lead to the Yukawa Lagrangian 
\be
\mathcal{L}_Y =
- y_1 \sum_\alpha \bar D_{L\alpha} \alpha_R \phi_\alpha
- y_2 \sum_\alpha \bar D_{L\alpha} \nu_{R\alpha}
\left( i \tau_2 \phi_0^\ast \right) 
+ \mbox{H.c.}
\label{Y}
\ee
This has,
remarkably,
only two Yukawa coupling constants.
The Lagrangian~(\ref{Y}) enjoys the accidental symmetry
\be
\mathbf{z}_\nu: \quad \phi_0 \to -\phi_0, \ 
\nu_{Re} \to - \nu_{Re}, \
\nu_{R \mu} \to - \nu_{R \mu}, \
\nu_{R \tau} \to - \nu_{R \tau}.
\ee

\paragraph{Charged-lepton masses:} The mass of the charged lepton $\alpha$
is $m_\alpha = \left| y_1 v_\alpha \right|$.
Therefore $m_e : m_\mu : m_\tau =
\left| v_e \right| : \left| v_\mu \right| : \left| v_ \tau \right|$.
In our model one cannot obtain a small $m_e$
by just tuning some Yukawa couplings---one really needs a small VEV $v_e$.
For instance,
even if there were no further Higgs doublets in the theory
and $v_\tau$ by itself alone saturated the relation
\be\label{vev-saturation}
v^2 \equiv \sum_k \left| v_k \right|^2 \le
\left( 174\, \mathrm{GeV} \right)^2,
\ee
one would still need $\left| y_1 \right| \sim 0.01$
because $m_\tau \approx 1.78\, \mathrm{GeV}$.
We would then have $\left| v_e \right| \sim 50\, \mathrm{MeV}$.
On the other hand,
we may assume that there are in the full theory
some extra Higgs doublets other than the four $\phi_k$, 
which give mass to the quarks---in particular,
for the top-quark mass a doublet with a large VEV is mandatory.
Then $\left| v_ \tau \right|$ may be significantly smaller
than $174\, \mathrm{GeV}$ and,
accordingly,
$\left| v_e \right|$ will be significantly smaller
than $50\, \mathrm{MeV}$ too;
in particular,
$\left| v_e \right| \sim m_e$ is possible.

\paragraph{Soft symmetry breaking:}
We assume that the $\mathbbm{Z}_3$ symmetry
and the three $\mathbf{z}_\alpha$ symmetries
are softly broken in the dimension-three
neutrino Majorana mass terms\footnote{If we want to predict
maximal atmospheric mixing,
then we must assume an $S_3$ instead of a $\mathbbm{Z}_3$ family symmetry and,
furthermore,
assume that the subgroup of $S_3$,
the $\mu$--$\tau$ interchange symmetry,
is preserved in the soft breaking~\cite{HPSmodel,3p,mu-tau}.}
\be
\mathcal{L}_M
= {\textstyle \frac{1}{2}}\, \sum_{\alpha, \beta}
\bar \nu_{R \alpha} \left( M_R \right)_{\alpha \beta} C
\bar \nu_{R \beta}^T + \mathrm{H.c.}
\ee
We furthermore assume that the symmetries $\mathbf{z}^h_\alpha$
are also softly broken,
now by the dimension-two terms $\phi_k^\dagger \phi_l$ ($k \neq l$)
in the scalar potential.
However,
we shall assume that \emph{the combined symmetry}
\be
\mathbf{z}^h_e \circ \mathbf{z}_\nu: \quad
\phi_0 \to - \phi_0, \
\phi_e \to - \phi_e, \
e_R \to - e_R, \
\nu_{Re} \to - \nu_{Re}, \
\nu_{R \mu} \to - \nu_{R \mu}, \
\nu_{R \tau} \to - \nu_{R \tau}
\ee
\emph{is broken only spontaneously}.
Then the scalar potential is
\ba
V &=& \sum_k \left[ \mu_k \phi_k^\dagger \phi_k
+ \lambda_k \left( \phi_k^\dagger \phi_k \right)^2 \right]
\no & &
+ \sum_{k \neq l} \left[
\lambda_{kl}\, \phi_k^\dagger \phi_k\, \phi_l^\dagger \phi_l
+ \lambda_{kl}^\prime\, \phi_k^\dagger \phi_l\, \phi_l^\dagger \phi_k
+ \lambda_{kl}^{\prime \prime} \left( \phi_k^\dagger \phi_l \right)^2 \right]
\no & &
+ \left( \mu_{0e}\, \phi_0^\dagger \phi_e
+ \mu_{\mu \tau}\, \phi_\mu^\dagger \phi_\tau
+ \mathrm{H.c.} \right).
\label{V}
\ea
The quartic couplings in $V$,
those with coefficients generically represented by the letter $\lambda$,
obey all the family symmetries of the model.
Because of the couplings
$\lambda_{kl}^{\prime \prime} \left( \phi_k^\dagger \phi_l \right)^2$,
the only $U(1)$ symmetry of $V$ is the one associated with weak hypercharge;
therefore,
there are no Goldstone bosons in the model.

\paragraph{Suppression of $v_0$:} If $\mu_0$ \emph{is positive},
then $v_0$ will induced by $v_e$
and by the first term in the last line of~(\ref{V}):
\be
v_0 \approx - v_e\, \frac{\mu_{0e}}{\mu_0}.
\ee
We envisage the possibility that $\left| y_2 \right|$ and,
possibly,
also $\left| y_1 \right|$ are of order 1,
because that would enhance scalar effects and the experimental testability
of our model, \textit{cf.}~sections~3 and~4 below.
Still,
it is possible,
as we mentioned earlier,
that $\left| y_1 \right| \sim 0.01$
and $\left| v_e \right| \sim 50\, \mathrm{MeV}$.
However,
even in that case $\left| v_0 \right|$
could easily be much smaller than $\left| v_e \right|$,
as we pondered in~\cite{nested}.
Indeed,
$\left| \mu_{0e} \right|$ could \emph{naturally}~\cite{tHooft} be small,
and there is no reason why $\mu_0$ should not be rather large,
maybe even of order TeV.
Then $\left| v_0 \right|$ might be
much smaller than $\left| v_e \right|$---this is the mechanism that
we envisaged in the introduction.
Thus:
\begin{itemize}
\item If there are in the theory some Higgs doublets beyond the four $\phi_k$,
the Yukawa coupling constant $y_1$ in~(\ref{Y}) may be of order one
and then $\left| v_e \right| \sim m_e$.
In that case, 
$\left| \mu_{0e} \right|$ and $\mu_0$
may be allowed to be of the same order of magnitude
and $\left| v_0 \right| \sim \left| v_e \right| \sim m_e$.
\item If there are only the four $\phi_k$,\footnote{In that case
we might envisage a scenario in which the VEVs of $\phi_\tau$
and $\phi_\mu$ would be responsible,
respectively,
for the masses of the up-type and down-type quarks.
Other possibilities may of course also be considered.}
then $y_1$ is much smaller than one
and $\left| v_e \right| \sim 100\, m_e$.
In that case,
we may naturally assume $\left| \mu_{0e} \right| \ll \mu_0$
because the theory acquires the extra symmetry $\mathbf{z}_\nu$
when $\mu_{0e} = 0$.
We might then still obtain $\left| v_0 \right| \sim m_e$.
\end{itemize}

\paragraph{Lepton mixing:}
The neutrino Dirac mass matrix is in our model
proportional to the unit matrix:
\be
M_D = y_2^\ast v_0 \mathbbm{1}.
\label{Dirac}
\ee
Therefore,
the lepton mixing matrix $U$,
which diagonalizes $\mnu$,
also diagonalizes $M_R$:
\ba
U^T \mnu U &=& \mathrm{diag} \left( m_1, m_2, m_3 \right),
\\
U^\dagger M_R U^\ast &=&
- \exp{\left[ 2 i \arg{\left( y_2^\ast v_0 \right)} \right]}\
\mathrm{diag} \left( M_1, M_2, M_3 \right),
\ea
the $m_j$ and $M_j$ ($j = 1, 2, 3$) being,
respectively,
the light- and heavy-neutrino masses.
Consequently,
there is a close relationship between the spectra
of the light and heavy neutrinos:
\be
M_j = \frac{\left| y_2 v_0 \right|^2}{m_j}.
\label{Mm}
\ee
Following our rationale,
we shall assume $\left| y_2 v_0 \right| \sim m_e$.

\section{Possible collider effects}

In our model we assume $m_D \sim m_e \sim 1\, \mathrm{MeV}$
while $m_R \sim 1\, \mathrm{TeV}$.
Otherwise our model is a normal seesaw model,
therefore in it the mixing between the light and the heavy neutrinos
is of order $m_D / m_R \sim 10^{-6}$.
This small mixing suppresses most possible signatures
of heavy right-handed neutrinos
that have been considered in the literature~\cite{literature}.
For instance,
the Drell--Yan production of a virtual $W$ boson
and its subsequent decay  ${W^\pm}^\ast \to \ell^\pm N_j$,
where $\ell^\pm$ is a charged lepton and
$N_j$ a heavy Majorana neutrino,
is negligible because it is suppressed
by the mixing between the light and heavy neutrinos. 

Suppose that the Higgs doublet $\phi_q$
(which may be one of our four doublets or an additional one)
couples to the quarks and,
in particular,
generates the top-quark mass.
Then we may envisage the Drell--Yan production
of a virtual $\phi_q^\pm$ at the LHC,
followed by the transition $\phi_q^\pm \to \phi_0^\pm$ 
and the decay $\phi_0^\pm \to \ell^\pm N_j$,
finally leading to heavy-neutrino production.
However,
since the VEV of $\phi_q^0$ is necessarily large
and the VEV of $\phi_0^0$ is very small,
the mixing $\phi_q^\pm$--$\phi_0^\pm$ will in general be small,
unless we invoke finetuning in the scalar potential.

\begin{figure}
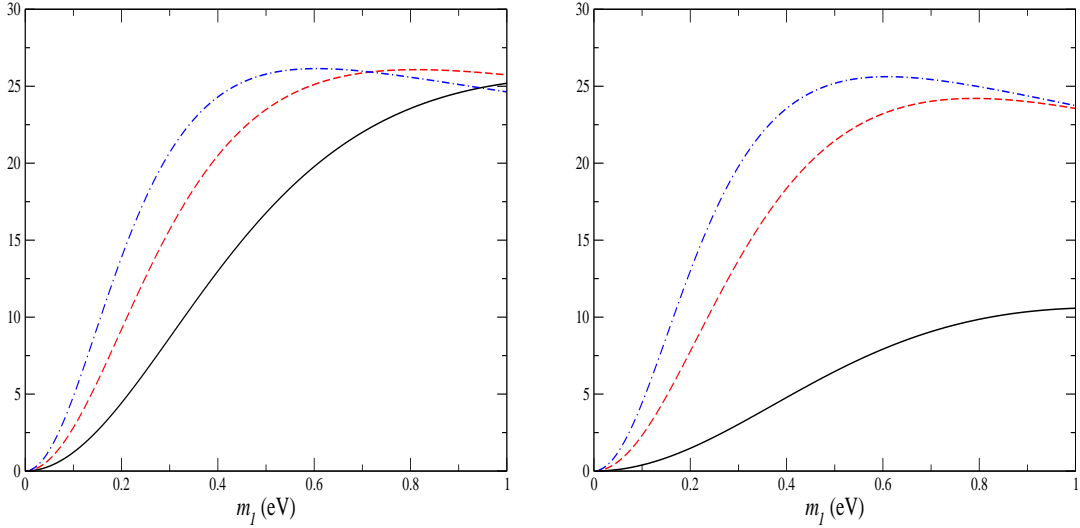

\begin{center}
\begin{tabular}{lcr}
\epsfig{file=fig120.eps,width=0.42\textwidth,height=70mm} 
& {} &
\epsfig{file=fig240.eps,width=0.42\textwidth,height=70mm}
\end{tabular}
\end{center}
\caption{$\left. \sigma \left( e^- e^- \to H^- H^- \right)
\, \right/ \sigma_ \mathrm{QED}$
as a function of $m_1$
when $\sqrt{s} = 500\, \mathrm{GeV}$ (full curves),
$750\, \mathrm{GeV}$ (dashed curves)
and $1\, \mathrm{TeV}$ (dashed-dotted curves).
The mass of $H^-$ is $m_H = 120\, \mathrm{GeV}$ in the left figure,
$m_H = 240\, \mathrm{GeV}$ in the right one.
We use equations~(\ref{Mm}),
(\ref{cs}) and~(\ref{f}),
$\left| y_2 \right| = 1$,
$\left| v_0 \right| = 511\, \mathrm{keV}$,
$U_{e1}^2 = 0.7$,
$U_{e2}^2 = 0.3$ and $m_2^2 = m_1^2 + 8 \times 10^{-5}\, \mathrm{eV}^2$.
\label{fig1}}
\end{figure}

In contrast to what happens at hadron colliders,
in an electron-electron collider
the interesting lepton-number-violating process
$e^- e^- \to H^- H^-$ might occur~\cite{soni1}.
This process is due to the Majorana nature of the heavy neutrinos
and is one of the processes, other than neutrinoless $\beta\beta$ decay,
via which it might be possible to probe
lepton-number violation~\cite{rodejohann}.
Let us suppose for simplification that $H^- \equiv \phi_0^-$,
the charged component of the scalar doublet $\phi_0$.
Then,
the relevant Yukawa Lagrangian for $e^- e^- \to H^- H^-$ is given by 
\be
\mathcal{L} \left( e^- e^- \to H^- H^- \right) =
y_2^\ast H^+ \sum_{j=1}^3 U_{ej}^\ast \bar N_j P_L e
+ \mbox{H.c.},
\ee
where $P_L$ is the negative-chirality projection matrix.
This leads to the total cross section (see~\cite{soni1} for a special case)
\be
\label{cs}
\sigma \left( e^- e^- \to H^- H^- \right) =
\frac{\left| y_2 \right|^4}{16 \pi s^2 \beta}
\sum_{j,k = 1}^3 \left( U_{ej} U_{ek}^\ast \right)^2 M_j M_k\,
f \left( \frac{w_j}{\beta}, \frac{w_k}{\beta} \right),
\ee
where $s$ is the square of the energy of the $e^- e^-$ system
in its centre-of-momentum reference frame,
$\beta = \left( 1 - 4 m_H^2 / s \right)^{1/2}$
($m_H$ is the mass of $H^-$)
and $w_j = 1 - 2 m_H^2 / s + 2 M_j^2 / s$.
The function $f$ is given by
\be
f \left( a, b \right) = \left\{ \begin{array}{lcl}
{\displaystyle \frac{1}{b^2 - a^2} \left(
b\, \ln{\left| \frac{a + 1}{a - 1} \right|}
- a\, \ln{\left| \frac{b + 1}{b - 1} \right|}
\right)}
& \Leftarrow & b \neq a,
\\
{\displaystyle \frac{1}{a^2 - 1} + \frac{1}{2 a}\,
\ln{\left| \frac{a + 1}{a - 1} \right|}}
& \Leftarrow & b = a.
\end{array} \right.
\label{f}
\ee
Notice that the cross section~(\ref{cs})
depends on the Majorana phases of the products
$\left( U_{ej} U_{ek}^\ast \right)^2$ ($j \neq k$).
In figure~\ref{fig1} 
we have plotted $\sigma \left( e^- e^- \to H^- H^- \right)$
as a function of the light-neutrino mass $m_1$
in a number of cases.
Following our rationale,
we have taken $\left| y_2 v_0 \right| = m_e$ in equation~(\ref{Mm}).
On the other hand,
in equation~(\ref{cs}) we have taken $\left| y_2 \right| = 1$,
bearing in mind that
$\sigma \left( e^- e^- \to H^- H^- \right)$ depends very strongly on
this Yukawa coupling.
As for the $U_{ej}$ matrix elements,
we have used the values $\left| U_{e1} \right|^2 = 0.7$,
$\left| U_{e2} \right|^2 = 0.3$
and $U_{e3} = 0$, which almost coincide
with the present best fit~\cite{fogli};
this has the advantage that then
the third-neutrino mass $m_3$ and the type of neutrino mass spectrum
become irrelevant---we only have to take into account
the experimental value of $m_2^2 - m_1^2$,
which we fixed at $8 \times 10^{-5}\, \mathrm{eV}^2$.
We have moreover assumed that the phase
of $\left( U_{e1} U_{e2}^\ast \right)^2$ is zero---this choice
maximizes $\sigma \left( e^- e^- \to H^- H^- \right)$.
In figure~\ref{fig1} this cross section is given
in units of $\sigma_\mathrm{QED} = 4 \pi \alpha^2 / \left( 3 s \right)$,
with $\alpha = 1 / 128$.

In the limit $M_j^2 \gg s, m_H^2$ for all $j = 1, 2, 3$,
one obtains
\be
f \left( \frac{w_j}{\beta}, \frac{w_k}{\beta} \right) \approx
\frac{2 \beta^2}{w_j w_k} \approx \frac{s^2 \beta^2}{2 M_j^2 M_k^2}.
\ee
Therefore,
in that limit
\be
\sigma \left( e^- e^- \to H^- H^- \right) \approx
\frac{\beta m_{\beta\beta}^2}{32 \pi \left| v_0 \right|^4},
\label{betasigma}
\ee
where $m_{\beta\beta} =
\left| \sum_j m_j \left( U_{ej} \right)^2 \right|$
is the effective mass measured in neutrinoless $\beta\beta$ decay.
The approximation~(\ref{betasigma}) indicates a close relationship and,
indeed,
an approximate proportionality
between $\sigma \left( e^- e^- \to H^- H^- \right)$ and $m_{\beta \beta}^2$.
Equation~(\ref{betasigma}) overestimates the cross section:
$\sigma \left( e^- e^- \to H^- H^- \right)$
is much smaller than indicated by that approximation
whenever $m_1 \gtrsim 0.5\, \mathrm{eV}$.
On the other hand,
for $m_1 \lesssim 0.1\, \mathrm{eV}$
the approximation~(\ref{betasigma}) becomes quite good,
but at the same time the cross section becomes small.
With the values of $U$ used in figure~\ref{fig1},
for $m_1 = 0.1\, \mathrm{eV}$,
$\sqrt{s} = 10^3\, \mathrm{GeV}$
and $m_H = 120\, \mathrm{GeV}$,
equation~(\ref{betasigma}) gives a cross section about 14\% too large;
for smaller $s$ or larger $m_H$ the discrepancy is smaller.
Note that a cross section of about $10^{-2} \times \sigma_\mathrm{QED}$
is still considered reasonable
for detection of $e^- e^- \to H^- H^-$
at an $e^- e^-$ collider~\cite{soni1}.

\section{Possible non-collider effects}
\label{non-col}

We next investigate whether
large Yukawa couplings $y_1$ and $y_2$ in equation~(\ref{Y})
might induce measurable effects in non-collider physics.

\subsection{The magnetic dipole moment of the muon}

A promising observable is $a_\mu$,
the anomalous magnetic moment of the muon.
There is a puzzling $3 \sigma$ discrepancy~\cite{RPP08}
\be
a_\mu^\mathrm{exp} - a_\mu^\mathrm{SM}
= 255(63)(49) \times 10^{-11}
\label{discrep}
\ee
between experiment and the SM prediction for that observable.
In our model there are contributions to $a_\mu$
from one-loop diagrams involving
either charged or neutral scalars.
We consider the latter firstly.
The fields $\phi_k^0$ are written
in terms of the physical neutral scalars $S_b^0$
($b = 2, \ldots, 8$)
as
\be
\phi_k^0 = v_k + \sum_{b=1}^8
\frac{\mathcal{V}_{kb} S_b^0}{\sqrt{2}}.
\ee
The scalar $S_1^0$ corresponds to the Goldstone boson
eaten by the gauge boson $Z^0$ and is,
therefore,
unphysical.
The complex $4 \times 8$ matrix $\mathcal{V}$ is such that
\be
\left( \begin{array}{c}
\mbox{Re}\, \mathcal{V} \\ \mbox{Im}\, \mathcal{V}
\end{array} \right)
\ee
is \emph{orthogonal};\footnote{For more details
on this notation see~\cite{precision}.}
therefore,
$\left| \mathcal{V}_{kb} \right|^2 \le 1$.
For $b \ge 2$,
let $m_b$ be the mass of $S_b^0$
and $\delta_b = m_\mu^2 / m_b^2 \ll 1$.
The contribution of the physical neutral scalars to $a_\mu$ is given by 
\ba
a_\mu \left( \phi^0 \right) &=&
\sum_{b=2}^8\, \frac{\delta_b}{16 \pi^2}\,
\int_0^1 \frac{\mathrm{d}x}{\delta_b x^2 - x + 1}
\left[ x^2\, \mathrm{Re}\, A_b
+ \left( x^2 - x^3 \right) \left| A_b \right| \right]
\\ &=&
\sum_{b=2}^8 \left\{ \frac{\delta_b}{16 \pi^2} \left[
\frac{\left| A_b \right|}{3}
- \left( \frac{3}{2} + \ln{\delta_b} \right) \mathrm{Re}\, A_b \right]
+ \mathrm{O} \left( \delta_b^2 \right) \right\},
\label{amu}
\ea
where $A_b = y_1^2 \mathcal{V}_{\mu b}^2$.
One sees that there is a term
$\delta_b \ln{\delta_b} \left/ \left( 16 \pi^2 \right) \right.$
which is $\sim 10^{-7}$ when $m_b \sim 100\, \mathrm{GeV}$.
Thus,
if one assumes $\left| A_b \right| \sim 1$
then $\left| a_\mu \left( \phi^0 \right) \right|$ might well be very large.

We proceed to analyze the diagrams involving charged scalars in the loop.
The three left-handed neutrinos $\nu_{L \alpha}$
and the three right-handed neutrinos $\nu_{R \alpha}$
are written in terms of the six physical Majorana neutrinos $\chi_i$
($i = 1, \ldots, 6$) as
\be
\nu_{L\alpha} = \sum_{i=1}^6 R_{\alpha i} P_L \chi_i, \quad
\nu_{R\alpha} = \sum_{i=1}^6 S_{\alpha i}^\ast P_R \chi_i,
\ee
where $P_{L,R}$ are the chirality projectors in Dirac space and $R$,
$S$ are $3 \times 6$ matrices.
The matrix
\be
\left( \begin{array}{c} R \\ S \end{array} \right)
\ee
is $6 \times 6$ unitary~\cite{GL02}.
The fields $\phi_k^+$ are written in terms of
the physical charged scalars $S_a^+$ ($a = 2, 3, 4$) as
\be
\phi_k^+ = \sum_{a=1}^4 \mathcal{U}_{ka} S_a^+.
\ee
The scalar $S_1^+$ corresponds to the Goldstone boson
eaten by the gauge boson $W^+$ and is,
therefore,
unphysical.
The complex $4 \times 4$ matrix $\mathcal{U}$ is unitary~\cite{precision}.

Let $m_i$ be the mass of the physical neutrino $\chi_i$ and,
for $a \ge 2$,
let $m_a$ be the mass of $S_a^+$.
The contribution of the diagrams with charged scalars to $a_\mu$ is given by 
\ba
a_\mu \left( \phi^+ \right) &=&
\frac{1}{16 \pi^2} \sum_{a=2}^4\, \sum_{i=1}^6\,
\int_0^1 \mathrm{d} x\, \frac{m_\mu^2}
{m_\mu^2 x^2 + \left( m_a^2 - m_\mu^2 - m_i^2 \right) x + m_i^2}
\no & & \times
\left[
\left( \left| y_1 R_{\mu i} \mathcal{U}_{\mu a} \right|^2
+ \left| y_2 S_{\mu i} \mathcal{U}_{0a} \right|^2 \right)
\left( x^3 - x^2 \right)
\right. \no & & \left.
+ 2\, \frac{m_i}{m_\mu}\, \mathrm{Re}
\left( y_1^\ast y_2^\ast R_{\mu i} S_{\mu i}
\mathcal{U}_{\mu a}^\ast \mathcal{U}_{0a} \right)
\left( x - x^2 \right) \right].
\label{amu+}
\ea
The three light neutrinos $\chi_{1,2,3}$ have masses much smaller than $m_\mu$,
hence these masses are negligible.
For those neutrinos the matrix elements $S_{\mu i} \sim 10^{-6}$
are also negligible.
One then has
\be
a_\mu \left( \phi^+ \right)_\mathrm{light\ neutrinos}
\approx
- \frac{\left| y_1 \right|^2}{96 \pi^2}
\sum_{a=2}^4 \delta_a \left| \mathcal{U}_{\mu a} \right|^2,
\ee
where $\delta_a = m_\mu^2 / m_a^2$.
For $m_a \sim 100\, \mathrm{GeV}$ one thus has
$a_\mu \left( \phi^+ \right)_\mathrm{light\ neutrinos}
\sim - 10^{-9} \left| y_1 \right|^2$,
which is not very significant.
The three heavy neutrinos $\chi_{4,5,6}$ have masses
comparable to those of the charged scalars.
For those neutrinos the $R_{\mu i} \sim 10^{-6}$.
One then has
\ba
a_\mu \left( \phi^+ \right)_\mathrm{heavy\ neutrinos}
&\approx&
\sum_{a=2}^4 \sum_{i=4}^6 \left\{
\frac{\left| y_2 S_{\mu i} \mathcal{U}_{0a} \right|^2}{16 \pi^2}\,
\frac{m_\mu^2}{m_i^2}\, f \left( \frac{m_a^2}{m_i^2} \right)
\right. \no & &
+ \frac{\mathrm{Re} \left( y_1^\ast y_2^\ast R_{\mu i} S_{\mu i}
\mathcal{U}_{\mu a}^\ast \mathcal{U}_{0a} \right)}{8 \pi^2}\,
\frac{m_\mu}{m_i}
\no & & \left.
\times \left[
\frac{1}{6} + \left( \frac{m_a^2}{m_i^2} - 1 \right)
f \left( \frac{m_a^2}{m_i^2} \right) \right],
\right\}
\ea
where
\ba
f \left( x \right) &=&
\frac{1}{3 \left( x - 1 \right)}
- \frac{x}{2 \left( x - 1 \right)^2}
+ \frac{x}{\left( x - 1 \right)^3}
- \frac{x \ln{x}}{\left( x - 1 \right)^4}
\\ &=&
- \frac{1}{3} + \left( - \frac{11}{6} - \ln{x} \right) x
+ \mathrm{O} \left( x^2 \right).
\ea
For Yukawa couplings of order 1,
and for heavy neutrinos
with masses of order $1\, \mathrm{TeV}$,
one obtains
$a_\mu \left( \phi^+ \right)_\mathrm{heavy\ neutrinos} \sim 10^{-10}$,
which is smaller than the experimental error in equation~(\ref{discrep}) and
hence irrelevant. 

So one concludes that
the largest non-standard contribution to $a_\mu$ in our model
is in principle $a_\mu \left( \phi^0 \right)$,
which is proportional to $\left| y_1 \right|^2$
and may be as large as $10^{-7}$ if $\left| y_1 \right|^2 \sim 1$.
One must invoke either a small Yukawa coupling $y_1$
or cancellations between the contributions of the various neutral scalars
for our model not to give too large a contribution to $a_\mu$.
One may also view~(\ref{amu})
as a possible way to explain the discrepancy~(\ref{discrep}).

\subsection{The decay $\mu^- \to e^- e^+ e^-$}

In our model, there are also 
lepton-flavour-changing processes
due to the soft breaking of the family lepton numbers~\cite{GL02}.
The process most likely to be observable is in general~\cite{GL02}
$\mu^- \to e^- e^+ e^-$,\footnote{The process $\mu^- \to e^- \gamma$
usually is more suppressed than $\mu^- \to e^- e^+ e^-$
in this type of models~\cite{GL02}.}
which is mediated by $\mu^- \to e^- {S_b^0}^\ast$ 
if the Yukawa couplings are large.
Using the formulas in~\cite{GL02} we obtain the approximate upper bound 
\be\label{br}
\mathrm{BR} \left( \mu^- \to e^- e^+ e^- \right) \lesssim 
10^{-5} \left| y_1 y_2 \right|^4
\left[ \sum_b \left| \mathcal{V}_{eb} \right|^2 
\frac{\left( 100\, \mathrm{GeV} \right)^2}{m_b^2} \right]^2
\left| F \right|^2,
\ee
where
\be
F = \sum_{i=1}^3 U_{e i} U_{\mu i}^\ast\, \ln{\frac{M_i^2}{{\bar\mu}^2}},
\ee
$\bar\mu$ being an arbitrary mass parameter
on which $F$ finally does not depend.
Clearly,
if $y_1$,
$y_2$ and $F$ are all of order one, 
then $\mathrm{BR} \left( \mu^- \to e^- e^+ e^- \right)$
will be much too large in our model,
since experimentally that branching ratio cannot be larger than
$1.0 \times 10^{-12}$ at 90\% confidence level~\cite{RPP08}.
One possibility to avoid this problem
is by making $\left| y_1 \right| \sim 0.01$,
which is possible as we have seen in section~\ref{model},
and would have the further advantage of also suppressing
$a_\mu \left( \phi^0 \right)$,
while keeping $y_2$ at order one
in order not to suppress $\sigma \left( e^- e^- \to H^- H^- \right)$.
Another possibility is to assume that
the mass spectrum of the light neutrinos is almost degenerate,
which in turn renders the mass spectrum of the heavy neutrinos
almost degenerate too.
With this condition and assuming $U_{e3}$ to be exactly zero,
we estimate
\be
\left| F \right| \simeq \left| U_{e2} U_{\mu 2} \right|
\frac{\deltasol}{m_1^2}.
\ee
Taking, for instance, $m_1 = 0.3\, \mathrm{eV}$,
one obtains $\left| F \right|^2 \sim 10^{-7}$,
which is sufficient to make equation~(\ref{br})
compatible with the experimental bound. 

Therefore,
our model may be compatible with the present experimental bound 
on $\mathrm{BR} \left( \mu^- \to e^- e^+ e^- \right)$, 
if the mass spectrum of the light neutrinos is almost degenerate 
or the Yukawa coupling $y_1$ is of order 0.01.
On the other hand,
the Yukawa coupling $y_2$ may perfectly well be of order one.

\section{Conclusions}

We have started in this Letter with the trivial observations that, 
in the seesaw mechanism,
a mass scale of $1\, \mathrm{MeV}$ in the neutrino Dirac mass matrix
corresponds to right-handed neutrinos in the TeV range,
and that the scale $1\, \mathrm{MeV}$ might be provided by the electron mass.
We have then constructed a simple model,
which extends the SM with three right-handed neutrino singlets
and four Higgs doublets,
in which that observation is realized.
Our model has the following properties:
\begin{enumerate}
\item Each charged lepton $\alpha$
has a corresponding Higgs doublet $\phi_\alpha$ which,
through its VEV $v_\alpha$,
generates the charged-lepton mass $m_\alpha = \left| y_1 v_\alpha \right|$.
\item The Dirac mass matrix $M_D$ 
of the neutrinos is generated by another Higgs doublet,
$\phi_0$,
whose VEV $v_0$ is induced by the VEV $v_e$
such that $v_0 \sim v_e$ or smaller,
hence $v_0 \propto m_e$.
\item In the appropriate weak basis,
the mass matrix of the charged leptons is diagonal
while $M_D$ is proportional to the unit matrix;
this yields the simple relation $M_j \propto \left. m_e^2\, \right/ m_j$
between the masses $M_j$ of the heavy Majorana neutrinos
and the masses $m_j$ of the light Majorana neutrinos. 
\item Moreover,
the diagonalization matrix of the light-neutrino mass matrix---which
is just the lepton mixing matrix---and
the diagonalization matrix of the heavy-neutrino mass matrix
are complex conjugate of each other.
\item The mixing between light and heavy neutrinos is small,
of order $10^{-6}$.
\end{enumerate}
The last property prevents heavy-neutrino production
in $pp$ and $p \bar p$ colliders.
However,
at an $e^- e^-$ collider
one might test the mechanism of the model
by searching for the process $e^- e^- \to H^- H^-$,
where $H^-$ is a charged scalar;
this process is somehow a high-energy analogue
of neutrinoless $\beta\beta$ decay.
The charged scalar would mainly decay
into a light neutrino and the electron,
if the heavy neutrinos are heavier than the scalar.
Thus the signal would amount to $e^- e^-$ plus missing momentum.
At an $e^+ e^-$ collider one should,
of course,
look for $e^+ e^- \to H^+ H^-$.

If the Yukawa coupling $y_1$ of our model is of order one,
then the anomalous magnetic moment of the muon may be too large,
and the branching ratio
of the flavour-changing process $\mu^- \to e^- e^+ e^-$
as well; in this case the suppression mechanisms discussed in
section~\ref{non-col} have to be invoked. 
The simplest way to make our model
compatible with experimental constraints is by choosing 
$\left| y_1 \right| \sim 0.01$. 
However,
this does not impede the process $e^- e^- \to H^- H^-$,
which proceeds through a different Yukawa coupling $y_2$.

\paragraph{Acknowledgements:}
We thank Juan Antonio Aguilar-Saavedra for valuable discussions.
Support for this work was provided by the European Union
through the network programme MRTN-CT-2006-035505.
The work of L.L.~is funded by the Portuguese
\textit{Funda\c c\~ao para a Ci\^encia e a Tecnologia}
through the project U777--Plurianual.

\end{document}